\documentstyle[epsf,preprint,aps]{revtex} \begin{document} \draft
\preprint{\it Optics Communications\/ \bf 142\rm, 308--314 
(1997)\hbox to 9cm{\hfill}} 
\title{
A method for reaching detection efficiencies
necessary for optical loophole--free Bell experiments}
\voffset=-30pt
\author{Mladen Pavi\v ci\'c}
\address{Max--Planck--AG Nichtklassische Strahlung,  
Humboldt University of Berlin, Rudower Chaussee 5, 
Geb\"aude 10.16, D--12484 Berlin, Germany\\ 
University of Zagreb, GF, Ka\v ci\'ceva 26, POB 217, HR--10001
Zagreb, Croatia\cite{mp}}


\maketitle \widetext \begin{abstract} 
A method for preparing a loophole--free four--photon Bell 
experiments which requires a detection efficiency of 67\%\ is 
proposed. It enables realistic detection efficiencies of 
75\%\ at a visibility of 85\%. Two type--II crystals each 
down convert one correlated photon pair and we entangle one photon 
from one pair with one photon from the other pair on a highly 
transparent beam splitter. The entanglement selects two other 
conjugate photons into a Bell state. Wide solid angles for the 
conjugate photons then enable us to collect close to 100\%\ of them. 
The cases when both photon pairs come from only one of the two 
crystals are successfully taken into account. Hardy's equalities 
are discussed.  
 
\end{abstract}
\bigskip\bigskip

\narrowtext

\section{Introduction} \label{sec:intro}

Two--photon interferometry using down converted photons has recently 
been extensively used for demonstrating violations of local
\cite{alley,om88,rt89,ci-e94,ps94,IPRA94,Ijosab95,Ipla-4} as well as 
non--local \cite{wzm91} hidden--variables theories. However, in spite 
of the recent improvement in the efficiencies of single--photon 
detectors (close to 80\%) \cite{kw-eb93}, all experiments carried out 
to date have had ten or more times fewer coincidence counts than 
singles counts and this, in effect, meant a detection efficiency under 
10\%. The reason for this lies in a directional uncertainty of signal 
with respect to idler photons. On every ten or more detected signal 
(idler) photons only one of conjugate idler (signal) photons finds 
its way to the other detector. (Detectors in two--photon experiments 
must have the same openings.) Therefore, all experiments carried 
out so far relied only on coincidence counts and 
herewith on additional assumptions---the \it no enhancement 
assumption\/ \rm was the most important one---which were considered 
very plausible. Then Santos devised \cite{San1,San2,San3} local 
hidden--variable models which violate not only the low detection 
loophole but also the \it no enhancement assumption\/\rm. 
These models, as well as improvements in techniques resulted in an 
interest into \it loophole--free\/ \rm experiments. In the past two 
years several sophisticated proposals appeared which rely on detailed 
elaborations of all experimental details. 
\cite{ci-e94,Ijosab95,Ipla-4,Fry,San4,jones} The last three 
proposals make use of maximal superpositions and require detection 
efficiency of at least 83\%\ \cite{83}, while the first three 
references make proposals for nonmaximal superpositions relying on
recent results \cite{eber93,Ipla,Inasa} which require only 67\%\
detection efficiency. In this paper we dispense with all 
supplementary assumptions by proposing a feasible selection 
method of doing a loophole--free Bell experiment which ideally 
requires only 67\%\ detection efficiency and 
can reach a realistic detection efficiency as high as 75\%. It is 
shown that this means a feasible conclusive experiment with a 
realistic visibility of 85\%. The method employs solid 
angles of signal and idler photons (in a type--II down conversion 
process) which differ five times from each other. This enables a 
detection of more than 90\%\ of conjugate photons. We also
consider a method of combining unpolarized independent photons into
spin correlated pairs by means of non--spin observables.  The physics
of such a preparation of spin states by means of non--spin observables
can be paralleled with the polarization correlation between unpolarized
photons discovered by Pavi\v ci\'c \cite{IPRA94} and formulated for
classical light by Paul and Wegmann \cite{paul94}. The main difference
is that in the latter experiments photons cross each other's paths
while in the present proposal they do not. In the end we compare
Hardy's equalities with the Bell inequalities.

\section{The experiment}
\label{sec:exp}

A schematic representation of the experiment is shown in
Fig.~\ref{exp}. An ultra--short laser beam (of frequency $\omega_p$), 
split by a beam splitter, simultaneously pumps up two type--II 
crystals. We assume that they are beta barium borate (BBO) crystals. 
In each type--II crystal the parametric down conversion produces 
intersecting cones of mutually perpendicularly polarized signal 
(extraordinary linearly polarized within \it e\/\rm--ray plane of the 
BBO) and idler (ordinary linearly polarized within \it o\/\rm--ray 
plane of the BBO) photons (of frequencies $\omega_e+\omega_o=\omega_p$). 
\cite{shih1} Signal and idler photons can be of the same 
frequencies ($\omega_e=\omega_o=\omega_p/2$) in which case the cones 
are tangent to each other. When we tilt the crystal so as to increase 
the angle between the pumping beam and the crystal axis of the BBO 
(increasing $\omega_o$ and decreasing $\omega_e$ slightly) the cones 
start to intersect each other [see inlet in Fig.~\ref{exp}]. Looking 
only at polarization, we see the photons at the intersections of 
the cones as entangled, because one cannot know which cone which 
photon comes from. \cite{gar} We then entangle one photon from one 
pair with one photon from the other pair by an interference of 
the fourth order at a beam splitter. Each successful entanglement 
(coincidence firing of detectors D1' and D2') selects (opens the 
computer gates for their counts) the other two conjugate photons 
into a Bell state. 

In a real experiment one first has to make photons at the cone 
intersections of each BBO indistinguishable, which means one has 
to compensate for the finite bandwidths and different group 
velocities inside the crystal, i.e., for transversal and 
longitudinal (\it e\/\rm--photon 
pulls ahead) walkoff effects. Half--wave plates (exchanging 
retardation of \it e\/ \rm and \it o\/ \rm photons) and quartz plates 
(being positive uniaxial crystal---BBO is negative) do the job. 
\cite{ci-e94,shih2,shih3,Kwiat} Then, by rotating the crystal, one can 
entangle the photons in a (non)maximal singlet--like or triplet--like 
state. \cite{Kwiat} In our proposal we assume both, crystals and 
plates, prepared so as to produce maximal singlet--like states. (It 
is interesting that starting with two maximal triplet--like states 
we arrive at the same final expressions for the probabilities;  
Cf.~Ref.~\cite{ps94}.) We also assume that the intensity 
of the laser pumping beam is reduced so that the probability 
of having more than two down converted singlets in chosen solid 
angles within the pumping time is 
negligible. \cite{ci-e94} We stress here that we choose a 
subpicosecond laser since without such an ultra--short pumping 
one would not be able to collect valid coincidence counts of D1' 
and D2' simply because there are no detectors which could react 
in a time short enough \cite{ps94} to confirm the intensity 
interference between two \it independent\/ \rm down converted 
photons (from two crystals) whose coherence time lies in a 
subpicosecond region. Two successive pumping can take place within 
several nanoseconds as determined by the lowest available detector 
resolution (recovering) time. For the feasibility of the experiment 
it is crucial that the probability of both photon pairs 
coming from only one of the two crystals can be made sufficiently 
small in comparison with the probability of photon pairs coming 
from both crystals by using more and more asymmetric beam splitter 
which at the same time lowers the required detection efficiency 
more and more towards 67\%. We show this at the end of this section.  

As we mentioned in Sec.~\ref{sec:intro}, the main detection 
efficiency problem in two photon interferometry is that signal 
and idler photons have 
to be in equal solid angles and that therefore less than 10\%\ 
of conjugate photons reach a detector. The present set--up enables 
us to use different solid angles for selecting photons (those which 
interfere at the beam splitter) and their conjugate photons (whose 
counts are passed by the gates). For the purpose, one has to 
evaluate the angular width of the conjugate photons once we know  
central directions (cone intersections) of both photons.    
One can show that the angular width of a conjugate photon depends 
on the frequencies of the pump, signal, and idler photons, 
on the band widths, on the pump, signal, and idler group refraction 
indexes, and on the directions of signal and idler photons with 
respect to the pumping beam, but for all combinations of these 
terms, the ratio of 1:5 between the solid angle of the 
photons we detect by detectors D1' or D2' and the solid angle 
centered around central direction of the conjugate photons ($ph$ 
in  Fig.~\ref{exp}) assures that over 90\%\ of conjugate photons 
will be found in the latter solid angle \cite{teich} and that the 
probability of detecting ``third party'' photons will be negligible. 
Let us now dwell on deriving our probabilities. 

We can have three input states, depending on whether the two 
pairs come from different crystals or both of them from only 
one of the crystals. The probabilities of the pairs being 
emitted in any of these three possible ways are equal.  
The singlets coming from the crystals are mutually 
independent and we therefore formally describe them by 
tensor products. The former one, coming from different 
crystals is given by  
\begin{eqnarray}
|\Psi\rangle={1\over\sqrt2}\bigl(|1_x\rangle_1|1_y\rangle_{1'}\>-
\>\,|1_y\rangle_1|1_x\rangle_{1'}\bigr)
\otimes{1\over\sqrt2}\bigl(|1_x\rangle_2|1_y\rangle_{2'}\>-
\>\,|1_y\rangle_2|1_x\rangle_{2'}\bigr)\,,\label{eq:4-state}
\end{eqnarray}
where $x$ corresponds to the \it e\/\rm--ray planes of the BBO's 
and $y$ to the \it o\/\rm--ray planes. The latter ones, coming from 
the same crystals are given by 
\begin{eqnarray}
|\Psi_{20}\rangle={1\over2}\bigl(|1_x\rangle_1|
1_y\rangle_{1'}-|1_y\rangle_1|1_x\rangle_{1'}\bigr)^2
|0\rangle_{2'}\,,\label{eq:2x0state}
\end{eqnarray}
\begin{eqnarray}
|\Psi_{02}\rangle={1\over2}\bigl(|1_x\rangle_2|1_y\rangle_{2'}-
|1_y\rangle_2|1_x\rangle_{2'}\bigr)^2
|0\rangle_{1'}\,.\label{eq:0x2state}
\end{eqnarray}

To obtain the four photon coincidence probabilities we cannot 
superpose these three input states upon one another because 
that would violate the principle of indistinguishability. To see 
this let us for the moment assume that our detection 
efficiency be ideal (100\%) and that the polarizers 
P1, P1$^\perp$, P2, and P2$^\perp$ be removed.  
Then, with the help of the responses of the detectors D1 and 
D2 we could tell $|\Psi\rangle$ (both D1 and D2 would 
fire), from $|\Psi_{02}\rangle$ (only D1 would fire) or 
from $|\Psi_{20}\rangle$ (only D2 would fire). 
If detections in reality had been ideal we would have 
used only $|\Psi\rangle$. Since they are not, we have to take 
$|\Psi_{02}\rangle$ and $|\Psi_{20}\rangle$ into account as 
well, but helpfully it turns out that the Bell inequality 
containing their corresponding counts (in addition to 
$|\Psi\rangle$ counts) is still violated. 
Therefore, we start with $|\Psi\rangle$, i.e., with 
two pairs coming out from different crystals, and discuss 
$|\Psi_{02}\rangle$ and $|\Psi_{20}\rangle$ later on.   
A multi--mode representation of the input state we give 
later on. 

The outgoing electric--field operators describing 
photons---we call them \it selector photons\/\rm---which 
pass through beam splitter BS, through polarizers P1' and P2' 
(oriented at angles $\theta_{1'}$ and $\theta_{2'}$, respectively), 
and are detected by detectors D1' and D2', read (see Fig.~\ref{exp})
\begin{eqnarray}
\hat E_{1'}&=&\left(\hat a_{1'x}t_x\cos\theta_{1'}+\hat
a_{1'y}t_y\sin\theta_{1'}\right)
e^{i\mbox{\scriptsize\bf k}_{1'}\cdot
\mbox{\scriptsize\bf r}_{1'}-
i\omega_1(t-t_{1'}-\tau_{1'})}\nonumber\\
&&+\ i\left(\hat a_{2'x}r_x\cos\theta_{1'}+\hat
a_{2'y}r_y\sin\theta_{1'}\right)
e^{i\tilde{\mbox{\scriptsize\bf k}}_{2'}\cdot
\mbox{\scriptsize\bf r}_{1'}-
i\omega_2(t-t_{2'}-\tau_{1'})}\,,\label{eq:E1'}
\end{eqnarray} 
\begin{eqnarray}
\hat E_{2'}&=&\left(\hat a_{2'x}t_x\cos\theta_{2'}+\hat
a_{2'y}t_y\sin\theta_{2'}\right)
e^{i\mbox{\scriptsize\bf k}_{2'}\cdot
\mbox{\scriptsize\bf r}_{2'}-
i\omega_2(t-t_{2'}-\tau_{2'})}\nonumber\\
&&+\ i\left(\hat a_{1'x}r_x\cos\theta_{2'}+\hat
a_{1'y}r_y\sin\theta_{2'}\right)
e^{i\tilde{\mbox{\scriptsize\bf k}}_{1'}\cdot
\mbox{\scriptsize\bf r}_{2'}-
i\omega_1(t-t_{1'}-\tau_{2'})}\,,\label{eq:E2'}
\end{eqnarray}
where $t_x^2$, $t_y^2$ are transmittances, $r_x^2$, $r_y^2$ are 
reflectances, $t_j'$ is time delay after which photon $j'$ reaches 
BS, $\tau_{j'}$ is time delay between BS and Dj',
$\omega_{j'}$ is the frequency of photon $j'$,  
$\mbox{\bf k}_{j'}$ is the wave vector of photon $j'$, and 
$\tilde{\mbox{\bf k}}_{j'}$ is the wave vector corresponding 
to $\mbox{\bf k}_{j'}$ after reflection at BS.  
The annihilation operators act as follows: 
${\hat a}_{jx}|1_{x}\rangle_{j'}=|0_{x}\rangle_{j'}$, \ 
${\hat a}_{jx}|0_{x}\rangle_{j'}=0$, $j'=1,2$. 

Operators describing photons---we call them \it Bell 
photons\/\rm---which pass through polarizers P1 and P2 (oriented 
at angles $\theta_1$ and $\theta_2$, respectively) and are 
detected by detectors D1 and D2 read 
\begin{eqnarray}
\hat E_1=
(\hat a_{1x}\cos\theta_1+\hat a_{1y}\sin\theta_1)
e^{-i\omega_1t_1}\,,\label{eq:E1}
\end{eqnarray}
\begin{eqnarray}
\hat E_2=
(\hat a_{2x}\cos\theta_2+\hat a_{2y}\sin\theta_2)
e^{-i\omega_2t_2}\,.\label{eq:E2}
\end{eqnarray}

The probability of detecting all four photons by detectors D1,
D2, D1', and D2' is thus
\begin{eqnarray}
P(\theta_{1'},\theta_{2'},\theta_1,\theta_2)
&=&\eta^2\langle\Psi|\hat E_{2'}^\dagger\hat E_{1'}^\dagger\hat E_2^
\dagger\hat E_1^\dagger
\hat E_1^{\phantom\dagger}\hat E_2^{\phantom\dagger}
\hat E_{1'}^{\phantom\dagger}\hat E_{2'}^{\phantom\dagger}|
\Psi\rangle\nonumber\\
&=&{\eta^2\over4}(A^2+B^2-2AB\cos\phi)\,,\label{eq:prob-4} 
\end{eqnarray} 
where $\eta$ is detection efficiency; 
$A=Q(t)_{11'}Q(t)_{22'}$ 
and $B=Q(r)_{12'}Q(r)_{21'}$; here 
$Q(q)_{ij}=q_x\sin\theta_i\cos\theta_j-q_y\cos\theta_i
\sin\theta_j$; $\phi=(\mbox{\boldmath$k$}_1-
\tilde{\mbox{\boldmath$k$}}_2)\cdot\mbox{\boldmath$r$}_1+
(\mbox{\boldmath$k$}_2-\tilde{\mbox{\boldmath$k$}}_1)
\cdot\mbox{\boldmath$r$}_2+(\omega_1-\omega_2)(\tau_{1'}-\tau_{2'})$. 

To obtain a realistic estimation of the above result we start 
with the multi--mode input states 
\begin{eqnarray}
|1\rangle_{1'}|1\rangle_{2'}=\int\int d\omega_1'
d\omega_1'\psi_1(\omega_1)\psi_2(\omega_2)|\omega_1\rangle_{1'}
|\omega_2\rangle_{2'}\,,\label{eq:int}
\end{eqnarray}
which we introduce into Eq.~(\ref{eq:4-state}); $\psi_i(\omega_i)$ 
($i=1,2$) are both peaked at $\omega={1\over2}\omega_p$: 
$\omega_i'=\omega-\omega_i$ ($i=1,2$). We can keep the singlet 
state description as given by Eq.~(\ref{eq:4-state}) because 
it has recently been proved by Keller and Rubin \cite{rubin}---as 
we briefly present below---that a subpicosecond pulse would give 
practically the same output as the continuous pumping beams provided 
a group velocity condition is matched. In doing so we 
rely on the experimental and theoretical results obtained by Kwiat 
\it et al.\/ \rm \cite{Kwiat}. We then make a 
Fourier decomposition of the electric--field operators 
[Eqs.~(\ref{eq:E1'}) and (\ref{eq:E2'})] and obtain the mean 
value for $P(\theta_{1'},\theta_{2'},\theta_1,\theta_2)$. 
By integrating the latter probability over $\tau_{1'}$, $\tau_{2'}$, 
$\omega_1'$, and $\omega_2'$ and using 
\begin{eqnarray}
{1\over T}\int_{-T/2}^{T/2}\cos(\omega\tau+a)\,d\tau=
{\sin(\omega T/2)\over\omega T/2}\cos a\,,
\qquad\qquad\qquad
\int_{-\infty}^{\infty}{\sin a\omega\over \omega}\sin b\omega 
\, d \omega=0
\,,\label{eq:int1}
\end{eqnarray}
\begin{eqnarray}
\int_{-\infty}^{\infty}{\sin a\omega\over \omega}\cos b\omega 
\, d \omega = 
\left\{
\begin{array}{ll}
\pi & \mbox{for $b<a$}\\
\pi/2 & \mbox{for $b=a$} \\
0 & \mbox{for $b>a$}\\
\end{array}
\right.\label{eq:int2} 
\end{eqnarray} 
we obtain  
\begin{eqnarray}
P(\theta_{1'},\theta_{2'},\theta_1,\theta_2)
={\eta^2\over4}(A^2+B^2-2ABv_e\cos\Phi)\,,\label{eq:prob-4e} 
\end{eqnarray} 
where 
\begin{eqnarray}
v_e={\int_{-T/2}^{T/2}f_1(\tau-\tau_1)f_2(\tau-\tau_2)\,d\tau\over 
\int_{-T/2}^{T/2}f_1^2(\tau-\tau_1)\,d\tau
+\int_{-T/2}^{T/2}f_2^2(\tau-\tau_2)\,d\tau}\,,\label{eq:ve}
\end{eqnarray} 
where $f_i(\tau)=\int_{-\infty}^{\infty}\psi_i(\omega)\cos 
\omega\tau\,d\omega$, ($i=0,1$), where $T$ is the detection time, 
and where $\Phi=2\pi(z_2-z_1)/L$; here $L$ is the spacing of the 
interference fringes,  $z_j$ are the coordinates of detectors D$j$'
along $\mbox{\boldmath$k$}_1-\tilde{\mbox{\boldmath$k$}}_2$ and 
$\mbox{\boldmath$k$}_2-\tilde{\mbox{\boldmath$k$}}_1$ (see Fig.~1 
in reference \cite{Ijosab95}); we dropped 
primes from $\tau_{1'}$ and $\tau_{2'}$ for simplicity. 
We see that $\Phi$ can be changed by moving the detectors 
transversally to the incident beams.     

By numerical calculation one can easily show that $v_e$ is not
susceptible to the variation of detection time $T$ provided
$|\tau_1-\tau_2|<|\omega_1-\omega_2|^{-1}$ (for $|\tau_1-
\tau_2|\ll|\omega_1-\omega_2|$ even when $T\gg |\omega_1-
\omega_2|^{-1}$). For $|\tau_1-\tau_2|\ll|\omega_1-\omega_2|^{-1}$ we 
have $v_e\rightarrow1$, i.e., the sharpest fringes, and for
$|\tau_1-\tau_2|\gg|\omega_1-\omega_2|^{-1}$ we have $v_e\rightarrow0$ 
and the fringes disappear. With the experimentally reachable frequency
passband $\Delta\omega$ of the order of magnitude of THz within a
single parametric down conversion with a continuous pumping beam
reaching the condition $|\tau_1-\tau_2|\ll|\Delta\omega|^{-1}$ is not a
problem because the time interval between the idler and signal photons
is of the order of femtoseconds. In our case, when dealing with two
simultaneous down conversions we have to resort to an ultra--short
pumping beam to satisfy the condition. Pulse pumping beam shorter than
1 ps would in general ``make it possible to distinguish pairs of
photons born at sufficiently different depths inside the crystal with
a consequent decrease in two photon interference'' as recently shown
by Keller and Rubin. \cite{rubin} This happens when the center of
momentum of signal and idler photons and the center of pump pulse does
not leave the crystal simultaneously.  When they do, i.e., when, by
choosing appropriate material conditions and pump frequency for a down
conversion within a type--II crystal, we make ``the inverse of the
pump group velocity equals the mean of the idler and signal inverse
group velocity'' \cite{rubin} and therewith we make the photons
indistinguishable again. Singlets appearing from such ``compensated''
crystal therefore keep their description given in Kwiat \it et al.\/
\rm \cite{Kwiat} and that is what we rely on in the afore--given
calculation.  

Another realistic detail of the experiment is that the
pinholes of detectors D1' and D2' are not points but have a certain
width $\Delta z$. Therefore, in order to obtain a realistic
probability we integrate Eq.~(\ref{eq:prob-4}) over $z_1$ and $z_2$
over $\Delta z$ to obtain \begin{eqnarray}
P(\theta_{1'},\theta_{2'},\theta_1,\theta_2)
&=&{\eta^2\over4}\int\limits_{z_1-{\Delta z\over 2}}
\limits^{z_1+{\Delta z\over 2}} \int\limits_{z_2-{\Delta
z\over2}}\limits^{z_2+{\Delta z\over 2}}
[A^2+B^2-2ABv_e\cos{2\pi(z_2-z_1)\over L}]dz_1dz_2\nonumber\\
&=&{\eta^2\over4}(A^2+B^2-v2AB\cos\Phi)\,,\label{eq:nu} \end{eqnarray}
where $v=v_e\bigl[\sin(\pi\Delta z/L)/(\pi\Delta z/L)\bigr]^2$ is the
\it visibility\/ \rm of the coincidence counting.

An analysis of Eq.~(\ref{eq:nu}) shows that triggering of 
D1' and D2' by selector photons means that their conjugate  
\rm Bell photons appear entangled in spite of the fact that they 
stem from two independent sources (two crystals) and that 
they do not interact in any way (e.g., they do not cross 
each other's paths). In general, Bell photons are only 
partially entangled as in the case of classical intensity 
interferometry. For special cases, however, one can achieve full 
quantum nonmaximal entanglement, i.e., one can obtain probability 
zero for certain orientations of polarizers P1 and P2. In order 
to obtain such an entangled state, which would at the same time 
enable a violation of the Bell inequality with only 67\%\ 
detection  efficiency, it is necessary to use an asymmetrical 
beam splitter, to orient polarizers P1' and P2', e.g., along 
$\theta_{1'}=90^\circ$ and $\theta_{2'}=0^\circ$, and to put 
detectors D1' and D2' in a symmetric position with respect to 
BS and with respect to the photons paths from the middle of the 
crystals so as to obtain $\Phi=0$. Eq.~(\ref{eq:prob-4}) then 
projects out the following nonmaximal singlet--like probability: 
\begin{eqnarray}
P(\theta_{1},\theta_{2})&=&
\eta^2s(\cos^2\theta_{1}\sin^2\theta_{2} - 
2v\rho\sin{\theta_{1}}\cos\theta_{1}
\sin{\theta_{2}}\cos\theta_{2}\cos\Phi+ 
\rho^2\sin^2\theta_{1}\cos^2\theta_{2})\,,\nonumber\\
&\equiv&\eta^2p(\theta_{1},\theta_{2})\label{eq:r-prob}
\end{eqnarray}
where we assumed the near normal incidence at BS so as to have 
$r_x^2=r_y^2=R$ and $t_x^2=t_y^2=T=1-R$, where we used 
$s\equiv T^2/(R^2+T^2)$, $\rho\equiv R/T$, and where we multiplied 
Eq.~(\ref{eq:nu}) by 4 for other three possible coincidence detections 
[i.e., for ($\theta_{1'}^{\phantom{\perp}}$,$\theta_{2'}^\perp$), 
($\theta_{1'}^\perp$,$\theta_{2'}^{\phantom{\perp}}$), and 
($\theta_{1'}^\perp$,$\theta_{2'}^\perp$) which we do not 
take into account because only 
($\theta_{1'}$,$\theta_{2'}$)--triggering opens the gates] and by 
$(R^2+T^2)^{-1}$ for photons emerging from the same side of BS 
(which also do not open the gates). 

The singles--probability of detecting a photon by D1  is
\begin{eqnarray}
P(\theta_1)= 
\eta s(\cos^2\theta_1+\rho^2\sin^2\theta_1)\equiv \eta p(\theta_1)\,
.\label{eq:one} 
\end{eqnarray}
Analogously, the singles--probability of detecting a photon by D2 is 
\begin{eqnarray}
P(\theta_2)= 
\eta s(\sin^2\theta_2+
\rho^2\cos^2\theta_2)\equiv \eta p(\theta_2)\,.\hfill\label{eq:one'} 
\end{eqnarray}

Introducing the above obtained probabilities into the 
Clauser--Horne \cite{cl-h} form of the Bell  inequality 
\begin{eqnarray}
B_{CH}\equiv P(\theta_1,\theta_2)-P(\theta_1,\theta_2')+
P(\theta_1',\theta_2')+P(\theta_1',\theta_2)-
P(\theta_1')-P(\theta_2)\leq0\,,\label{eq:loop} 
\end{eqnarray} 
we obtain the following minimal efficiency for its violation 
\begin{eqnarray}   
\eta={p(\theta_1')+p(\theta_2)\over 
p(\theta_1,\theta_2)-p(\theta_1,\theta_2')
+p(\theta_1',\theta_2')+p(\theta_1',\theta_2)}
\,.\hfill\label{eq:eta} 
\end{eqnarray}
We stress here that the probabilities in Eq.~(\ref{eq:loop})
are proper probabilities---not the ratios of coincidence 
counts as in the experiments carried out so far. For example, 
$P(\theta_2)=\eta p(\theta_2)$ is a total number of counts 
detected by detector D2 with the polarizer P2 oriented along 
$\theta_2$---it is \it not\/ \rm either $\eta^2p(\infty,a_2)$ or 
$\eta^2p(\infty,a_2)/p(\infty,\infty)$. 

This efficiency is a function of visibility $v$ and by looking at 
Eqs.~(\ref{eq:r-prob}), (\ref{eq:one}), and (\ref{eq:one'})  
we see that for each particular $v$ a different set of angles 
should minimize it. A computer optimization of angles---presented in 
Fig.~\ref{vis}---shows that the lower the reflectivity is, the lower 
is the minimal detection efficiency. Also, we see a  rather 
unexpected property that a low visibility does not have a significant 
impact on the violation of the Bell inequality. For example, with 
70\%\ visibility and 0.2 reflectivity of the beam splitter we obtain 
a violation of Eqs.~(\ref{eq:loop}) with a lower detection efficiency 
than with 100\%\ visibility and 0.5 ($\rho=1$) reflectivity. 

In Ref.~\cite{Ijosab95} we have shown that one can select 
fully quantum entangled Bell photons even without polarizers 
P1' and P2'; i.e., whenever unpolarized selector photons 
trigger detectors D1' and D2' they open the gates for maximally 
entangled singlet--like state of Bell photons. Now, it is of 
interest to find out whether we can use such non--polarization 
preparation to prepare full non--maximal polarization--entangled 
states. To this aim, we calculate
\begin{eqnarray}
P_{\infty}(\theta_1,\theta_2)=
P(\theta_{1'},\theta_{2'},\theta_1,\theta_2)+
P(\theta_{1'}^\perp,\theta_{2'},\theta_1,\theta_2)+
P(\theta_{1'},\theta_{2'}^\perp,\theta_1,\theta_2)+
P(\theta_{1'}^\perp,\theta_{2'}^\perp,\theta_1,\theta_2)
\label{eq:infty} 
\end{eqnarray} 
where we obtain the last three probabilities by analogy 
with the first one [Eq.~(\ref{eq:prob-4})]; e.g., in order to obtain  
$P(\theta_{1'}^\perp,\theta_{2'},\theta_1,\theta_2)$, we 
introduce $E_{2'}^\perp$ instead of $E_{2'}$ into Eq.~(\ref{eq:prob-4}), 
where we get $E_{2'}^\perp$ from Eq.~(\ref{eq:E2'}) upon substituting 
$-\sin\theta_{2'}$ for $\cos\theta_{2'}$ and  $\cos\theta_{2'}$ for 
$\sin\theta_{2'}$. Eq.~(\ref{eq:infty}) yields  
\begin{eqnarray}
P_{\infty}(\theta_1,\theta_2)=
\eta^2{(1-2r_x^2t_x^2)\sin^2\theta_{1}\sin^2\theta_{2} + 
(1-2r_y^2t_y^2)\cos^2\theta_{1}\cos^2\theta_{2}+S-2vW\cos\Phi\over
2(1-2r_x^2t_x^2-2r_y^2t_y^2+t_x^2t_y^2+r_x^2r_y^2)}
\label{eq:infty-1}
\end{eqnarray}
where 
\begin{eqnarray}
S=(t_x^2t_y^2+r_x^2r_y^2)(\sin^2\theta_{1}\cos^2\theta_{2}+
\cos^2\theta_{1}\sin^2\theta_{2})\,, 
\label{eq:S1}
\end{eqnarray}
\begin{eqnarray}
W=(t_xr_x\sin\theta_{1}\sin\theta_{2}+
t_yr_y\cos\theta_{1}\cos\theta_{2})^2\,. 
\label{eq:S2}
\end{eqnarray}
A computer calculation shows that this probability can violate 
the Bell inequalities only for a detection efficiency 83\%\ or 
higher. It also shows that the probability cannot be used for 
obtaining Hardy's equalities \cite{hardy}. On the other hand, 
an analysis of Eq.~(\ref{eq:infty-1}) shows that the only 
way to obtain a non--partial, i.e., full quantum (non--classical) 
entanglement is to use a symmetric beam splitter ($r_x^2=r_y^2=1/2$) 
and a symmetric position of detectors D1' and D2' with respect to 
BS and with respect to the photons paths from the middle of the 
crystals so as to obtain $\Phi=0$. Under these conditions 
Eq.~(\ref{eq:infty-1}) yields $P_{\infty}(\theta_1,\theta_2)=
{1\over 2}\sin^2(\theta_{1}-\theta_{2})$, 
i.e., a maximal singlet--like state. Thus, by means of non--spin 
preparation we can prepare only ``symmetric'' (maximal) 
non--classical spin correlated states.  

In the end we show that other down conversions which may 
occur in the crystals and enter our statistics do not significantly 
influence the obtained probabilities. The probability of both 
photon pairs coming from only one of the two crystals 
and the probability of their coming from both
crystals are of course equal, but for $\rho$ close to 0 the 
influence of photon pairs coming from only one of the two crystals 
can be made small enough for a conclusive Bell experiment. Let us 
see this in detail. 

Choosing $\theta_{1'}=90^\circ$, $\theta_{2'}=0^\circ$,  
$\Phi=0$,  and rewriting the electric--field 
operators [Eqs.~(\ref{eq:E1'}) and (\ref{eq:E2'})] accordingly, we 
obtain the following probabilities of detecting the  
``intruder'' counts (corresponding to both photons coming from the 
same crystal and being both detected by D1 and D2, respectively) while 
collecting the singles--probabilities [Eqs.~(\ref{eq:one}) and 
(\ref{eq:one'})]
\begin{eqnarray}
P_{20}(90^\circ,0^\circ,\theta_1,-)=P_{20}(\theta_1)
=\eta s \rho (1+v)\sin^2(2\theta_{1})\,,\label{eq:20-prob}
\end{eqnarray}
\begin{eqnarray}
P_{02}(90^\circ,0^\circ,-,\theta_2)=P_{02}(\theta_2)
=\eta s \rho (1+v)\sin^2(2\theta_{2})\,.\label{eq:02-prob}
\end{eqnarray}
We could dispense with these counts only if detectors D1 and D2 
could tell one photon from two. It is therefore important to see 
whether the Bell inequality Eqs.~(\ref{eq:loop}) is still violated 
when we have them in our statistics. In order to include them into 
the Bell inequality we should add them to the singles--probabilities 
given by Eqs.~(\ref{eq:one}) and (\ref{eq:one'}). By comparing 
$P_{20}(\theta_1)$ and $P_{02}(\theta_2)$ with $P(\theta_1)$ 
and $P(\theta_2)$, respectively, we see that for the angles 
close to $\pi\over2$ and $\pi$, for which the asymmetrical states 
violate the Bell inequality, the following inequalities hold: 
$P_{20}(\theta_1)\ll P(\theta_1)$ and 
$P_{02}(\theta_2)\ll P(\theta_1)$.  For example, for 
$\rho=0.1$, $\eta=0.75$, $v=0.9$, $\theta_1=104^\circ$, $\theta_1'=
89^\circ$, $\theta_2=181^\circ$, and $\theta_2'=161^\circ$ 
we obtain the violation $B_{CH}>0$. For the same parameters 
we also obtain $B_{CH}-P_{20}-P_{02}>0$. However, this reduces the 
value of $B_{CH}$ for which the Bell inequality is violated by 2/3. 
On the other hand, we have to use birefringent polarizers P1 and P2 
to be able to discard counts which fire both D1 and D1$^\perp$ when 
collecting data for the singles--probability $P(\theta_1)$  by 
D1 and those which fire both D2 and D2$^\perp$ when 
collecting data for the singles--probability $P(\theta_2)$  by 
D2. Therefore, in doing a real experiment we should better
split unwanted two--photon wave packets across  
additional polarized beam splitters \cite{rt89} or, even better,  
by applying \it photon chopping\/ \rm \cite{chop} when 
collecting counts for singles--probabilities. 
We stress here that by this method we do not affect 
the conclusiveness of the Bell experiment but only pick out \it 
valid\/ \rm Bell pairs from all those 
ones already selected by the D1'--D2' coincidence gates. 
That is, we do not discard any counts corresponding to firing of D1 
and/or D2 by photons coming from different crystals. 

\section{Conclusion}
\label{sec:concl}
Our elaboration shows that the proposed loophole--free Bell  
experiment which selects two out of four photons into 
nonmaximal singlet--like states can be carried out with the 
present technique. The proposal makes use of an asymmetrical 
preparation of two \it input\/ \rm photon singlets generated by 
two nonlinear type--II crystals. The asymmetry consists in the 
fact that we first let one photon from the first singlet interfere 
in the fourth order with one photon from the other singlet at a 
highly transparent beam splitter. Coincidental detections of the 
photons interfering at the beam splitter (we call them \it 
selector photons\/\rm) open gates for a 
selection of the remaining two conjugate photons, one from each 
singlet, into a new correlated state: nonmaximal \it Bell singlet\/\rm.  
In other words, since no coincidence detection between signal and idler
photons of the input singlets is needed we can use several times 
wider solid angles for the \it Bell photons\/ \rm than for 
the \it selector photons\/\rm. With five times wider solid angle 
(determined by pinholes \it ph\/ \rm in Fig.~\ref{exp}) we collect 
practically all Bell companions of those \it selector 
photons\/ \rm which trigger detectors D1' and D2' in coincidence. 
In this way we eliminate the main cause of the low detection 
in all two--photon experiments so far: loosing of the conjugate 
photons (in most cases they ``miss'' the detector opening). 
An apparent draw--back to our set--up is that the 
probabilities of two pairs coming from both crystals and of 
both pairs coming from only one of the crystals are equal. 
However, the above calculations show that for reflectivity 0.1,  
realizable visibility of 85--90\%, and achievable detector 
efficiency of 75\%\ \cite{ci-e94,kw-eb93} the Bell 
inequality is violated even when the counts corresponding to 
photons emerging from only one of the two crystals are included  
into the statistics by which the inequality is fed.   

We should mention here that although 67\%\ efficiency result for 
Hardy's equalities has been obtained recently as well 
\cite{gar2,china} their low marginal violations (of maximal value 
0.09 as opposed to 0.41 for the Bell inequalities) make a 
loophole--free ``Hardy experiment'' more demanding. However, it 
would be worth trying to collect data for it within the proposed 
set--up because of its conceptual clarity and because our results 
add to the physics of the Hardy experiment. In particular, 
an analysis of Eq.~(\ref{eq:infty-1}) shows that Hardy's equalities, 
as opposed to the Bell inequalities, cannot be formulated for a 
system which is not fully non--classical.  Thus, our set--up 
reveals nonlocality of quantum systems as a property of selection 
of their subsystems and Hardy's equalities as a test (ideally, some 
detectors should always react and some never) of whether the system 
is fully quantum or not. It may turn out that quantum nonlocality is 
only operationally defined in the same way in which quantum 
phase might turn out to be only operationally defined. 
\cite{mandel} On the other hand, since Hardy's equalities reach 
their maximal violation for $R=0.32$ and not for $R=0.5$, it might 
turn out the unwanted effect of both photon pairs coming from the 
same crystal on the marginal probabilities can be compensated 
sufficiently well to make the experiment feasible. 

In the end we mention that the set--up may find its application in 
quantum cryptography and quantum computation for its property to 
deliver Einstein--Podolsky--Rosen 
singlets \cite{cryp} whose ``coherence ... [is] retained over 
considerable distances and for long times'' \cite{sten}; 
actually, our \it Bell singlets\/ \rm stay coherent for ever, i.e., 
until we make use of them and collapse them.      

\bigskip
\bigskip
\parindent=0pt
\it Note.\/ \rm\ Some preliminary results to these paper have 
been presented within an invited talk at the \sl Adriatico 
Research Conference on\/ \it Quantum Interferometry II\/\rm,  
held in Trieste, Italy, 4--8 March 1996. \cite{Trst} 

\acknowledgments
The author thanks to Harry Paul, Humboldt--Universit\"at zu 
Berlin for many valuable discussions. He also acknowledges 
supports of the Max--Plack--Geselschaft, Germany.

\begin{figure}
\caption{Lay--out of the proposed experiment. Beam splitter BS and
detectors D1' and  D2', and their counters [which open the gates 
for Bell photon singlets] serve as an event--ready 
\protect\cite{cl-h} selection device. \it cp\/\rm's 
(compensation plates) represent half--wave plates and quartz plates. 
Inlet in the upper right corner shows intersecting cones of 
down converted photons emerging from type--II crystals. \it s\/ \rm 
represents solid angles (along the intersection of the cones) 
determined by the pinholes of detectors D1' or D2'.}
\label{exp} 
\end{figure}

\begin{figure}
\caption{Minimal detection efficiencies $\eta$ necessary for a 
violation of the Bell--Clauser--Horne inequality as functions of 
visibility $v$ and of $\rho=R/(1-R)$, where $R$ is the reflectivity 
of the beam splitter.} 
\label{vis} 
\end{figure}

\vfill\eject 

\epsffile{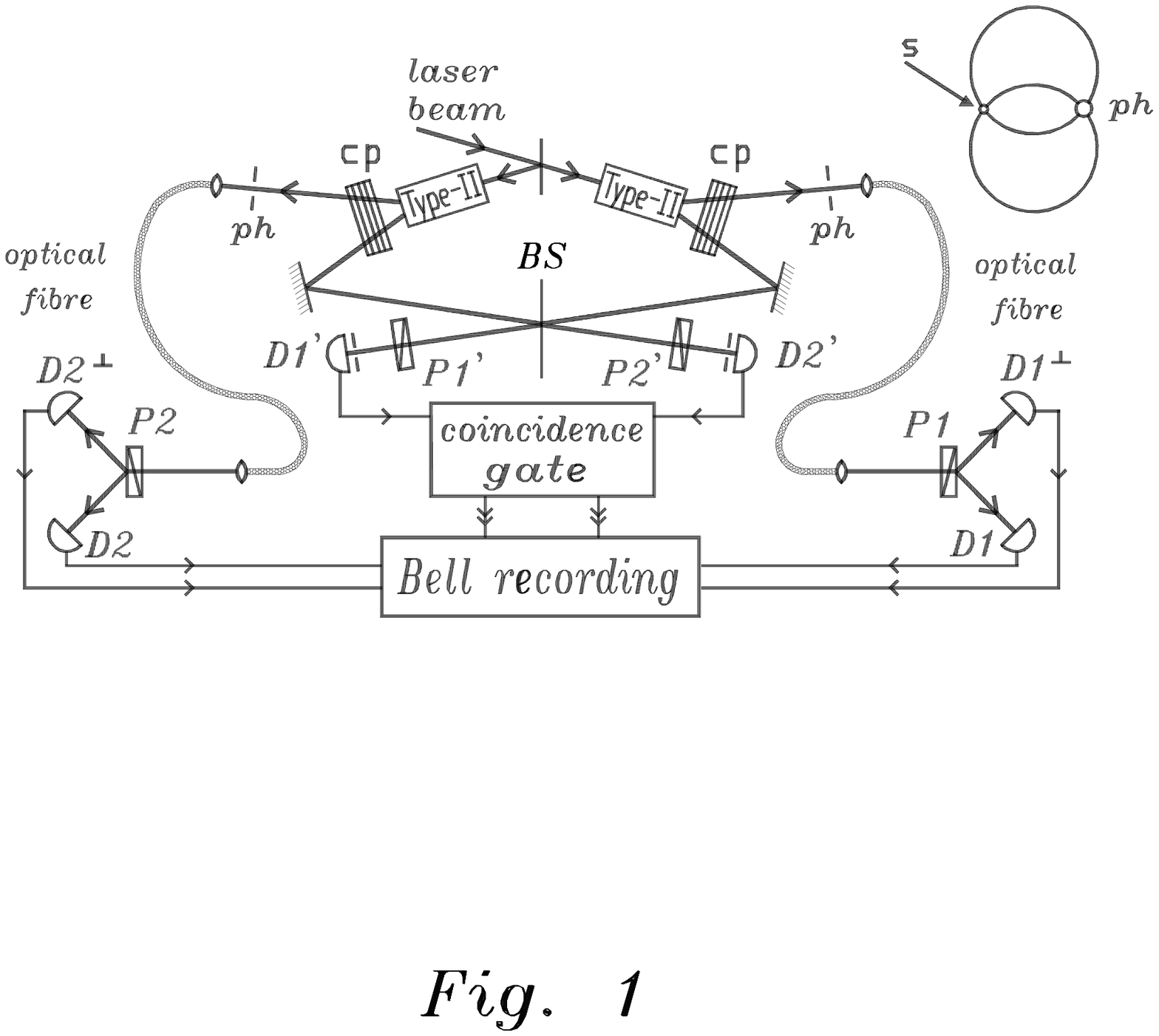}

\vfill\eject 

\epsffile{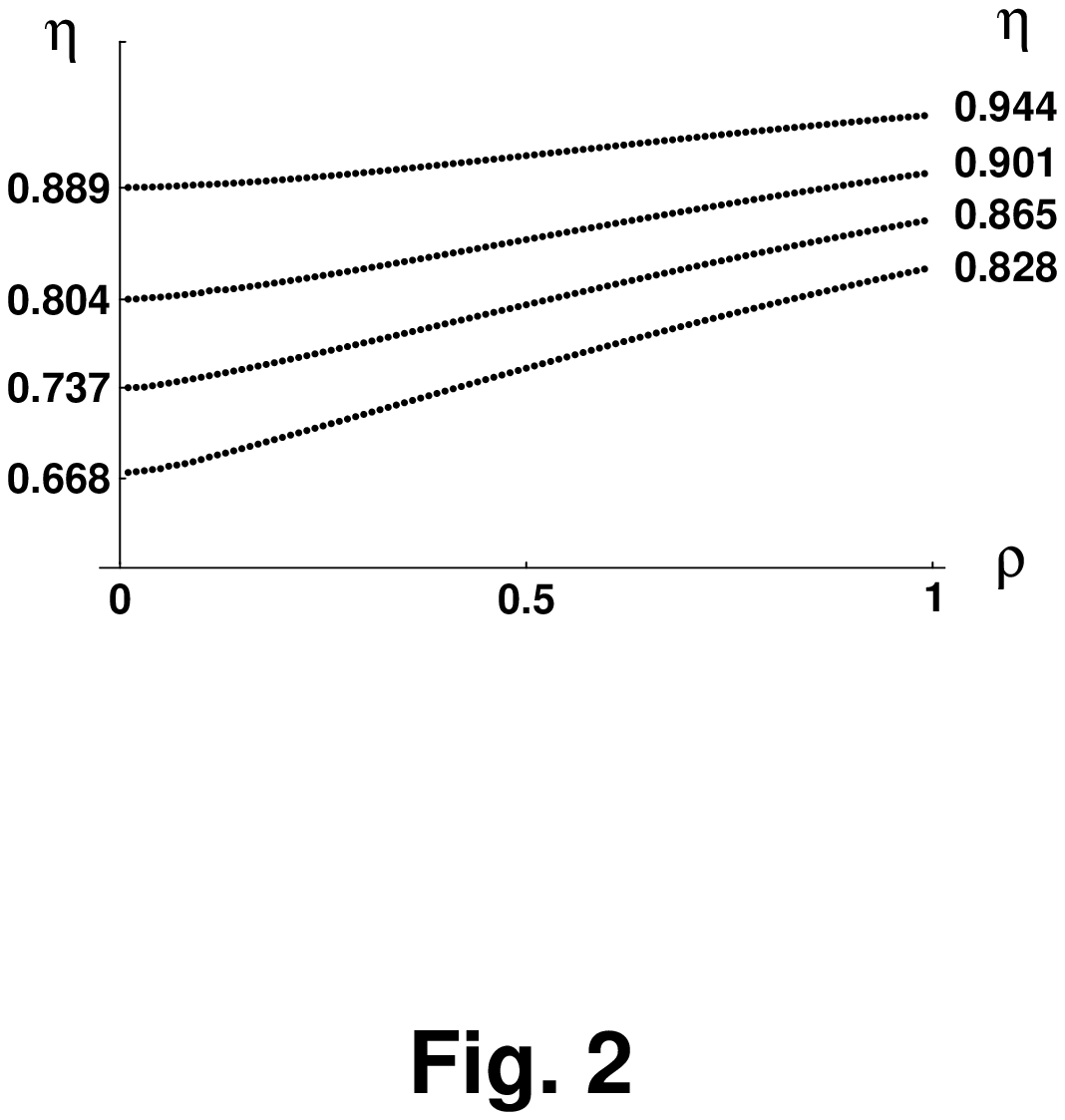}

\end{document}